\documentclass[twocolumn,english,showpacs]{revtex4}
\usepackage{babel,amsmath,amssymb,dcolumn}
\usepackage[dvips]{graphics}

\begin{document}

\title{Late time tails of the massive vector field in a black hole background}

\author{R.A. Konoplya}
\email{konoplya@fma.if.usp.br}
\affiliation{Instituto de F\'{\i}sica, Universidade de S\~{a}o Paulo \\
C.P. 66318, 05315-970, S\~{a}o Paulo-SP, Brazil}

\author{A. Zhidenko}
\email{zhidenko@fma.if.usp.br}
\affiliation{Instituto de F\'{\i}sica, Universidade de S\~{a}o Paulo \\
C.P. 66318, 05315-970, S\~{a}o Paulo-SP, Brazil}

\author{C. Molina}
\email{cmolina@usp.br}
\affiliation{Escola de Artes, Ci\^{e}ncias e Humanidades, Universidade de
  S\~{a}o Paulo\\ Av. Arlindo Bettio 1000, CEP 03828-000, S\~{a}o
  Paulo-SP, Brazil}


\begin{abstract}
We investigate the late-time behavior of the massive vector field in
the background of the Schwarzschild and  Schwarzschild-de Sitter black
holes. For  Schwarzschild black hole, at intermediately late times the
massive vector  field is represented by three functions with different
decay law $\Psi_{0} \sim t^{-(\ell + 3/2)} \sin{m t}$,  $\Psi_{1} \sim
t^{-(\ell + 5/2)} \sin{m t}$, $\Psi_{2} \sim t^{-(\ell + 1/2)}
\sin{m t}$, while at asymptotically late times the decay law $\Psi
\sim t^{-5/6} \sin{(m  t)}$ is universal, and does
not depend on the multipole number $\ell$.  Together with previous
study of massive scalar and Dirac fields where the same asymptotically
late-time decay law was found, it means, that the asymptotically
late-time decay law $\sim t^{-5/6} \sin{(m  t)}$ \emph{does not
  depend} also \emph{on the spin} of the field under
consideration. For Schwarzschild-de Sitter black holes it is observed
two different regimes in the late-time decay of perturbations:
non-oscillatory exponential damping for small values of $m$ and
oscillatory quasinormal mode decay for high enough $m$. Numerical and
analytical results are found for these quasinormal frequencies.
\end{abstract}

\maketitle

\section{Introduction}

Black hole's response to an external perturbation has been a subject of
active investigations for recent ten years \cite{Kokkotas-99}. Now it is a well
known fact that in the response signal of a black hole, the so-called
oscillatory \emph{quasinormal frequencies} dominate, which do not depend on
excitations of the quasinormal modes, but only on a black hole
parameters, thereby giving us the complete information about the
geometry of a black hole in the fully non-linear general relativity
theory \cite{kuchamala}.

As was shown for the first time by R. Price,
the regime of quasinormal oscillations turns into decay with inverse
power dependence on time for a  Schwarzschild black hole \cite{Price1}.
Price showed that perturbations of the massless scalar and
gravitational fields decay as $t^{-(2 \ell + 3)}$ at
asymptotically late times $t \rightarrow \infty$. Bicak found that
scalar massless field in the Reissner-Nordstr\"{o}m background decays as
$t^{-(2 \ell + 2)}$ for $|Q| < M$, and  as $t^{-(\ell + 2)}$
for  $|Q| = M$ \cite{Bicak}. The massless scalar field perturbations
along null infinity and along future event horizon has the decay law
$u^{-\ell + 2}$ and $v^{-\ell + 3}$ respectively, where $u$ and
$v$ are outgoing and ingoing Eddington-Filkenstein coordinates
\cite{Pullin}. In \cite{Hod-Piran} it was shown that the charged
scalar perturbations decay slower then a neutral ones at
asymptotically late times, while, on the  contrary, at the stage of
quasinormal ringing, the neutral perturbations decay slower
\cite{konoplya2002_2,konoplyaPLB}. Also, for Schwarzschild-de Sitter
and  Reissner-Nordstr\"{o}m-de Sitter black holes, instead of power-law tails,
the exponential tails were found \cite{sdstail}.

The investigation of late time decay for massless fields was continued
for $D$-dimensional black holes. Thus, tails of massless scalar,
vector and gravitational fields for higher-dimensional Schwarzschild
black holes have the decay law $t^{- (2 \ell + D -2)}$ for odd $D >
3$, and for even ($D > 4$) dimensions, the power law
decay is $t^{- (2 \ell + 3 D - 8)}$ \cite{Cardoso-tail}. The late-time
behavior for $D$-dimensional Gauss-Bonnet, Gauss-Bonnet-de Sitter and
Gauss-Bonnet-anti-de Sitter black holes was elaborated in
\cite{AKM}. Late time tails were also observed perturbations around
wormholes between the branes in a two brane Randall-Sundrum model
\cite{wormhole}. In addition, the late-time tails can be a tool
to probe the extra dimensions both at the ringing stage
\cite{higherD} and at the stage of late-time tails \cite{Barvinsky}.

The late time behavior of massive fields is qualitatively different
from massless ones: at late times the decay profile is \emph{oscillatory}
inverse power tail. For a massive scalar field with mass $m$ in the
background of the Schwarzschild black hole with mass $M$, the
perturbation decays as $t^{-(\ell +3/2)} \sin{m t}$ at
intermediate late times $mM < mt < 1/(m M)^{2}$ \cite{HodPiran1}, and as
$t^{-5/6} \sin{m t}$ at asymptotically late times
\cite{koyama1,koyama2}. The same $\sim t^{-5/6}$ behavior
was found for the massive scalar field perturbations of the Kerr black
hole \cite{burko} (but with frequency which is slowly dependent on
time  $\Psi \sim t^{-5/6} \sin{(\omega (t) \times t)}$), of the
dilaton black hole \cite{rogatko1}, and also, for massive Dirac
perturbations of the  Schwarzschild black hole \cite{jing}. Recently,
it has been found in \cite{rogatko2} that intermediate late time decay
of massive scalar field in $D$-dimensional Schwarzschild black hole is
proportional to $t^{\ell + D/2 - 1/2}$.

On the other hand, the late time tails of fields corresponding to
massive bosons have not been studied before, even for the simplest and
a most interesting background geometry, namely, for the Schwarzschild
black hole. In the present paper we shall investigate the late time behavior of
massive vector field, which shows quite non-trivial behavior also at the
stage of quasinormal ringing \cite{konoplya_sp1}. The massive vector
field obeys the Proca equation which can be reduced to a single
wave-like equation for spherically symmetrical perturbations
\cite{konoplya_sp1}. Therefore we are able to investigate the tail
behavior for spherically symmetric perturbations both numerically and
analytically. Yet, for higher multipoles $\ell > 0$ the Proca equations
cannot be decoupled. Fortunately, in this case we also can investigate the
tail behavior, because we need only the asymptotic form of the wave
equations at large $r$, rather than its exact form.

We found that at intermediately late times the three types of
perturbations of the vector field decay with three different decay laws,
while, at infinitely late times, these three types of decay approach
a single {\it universal} decay law, which is independent of multipole
number $\ell$, and is the same for massive fields of other spin.
Late time behavior for asymptotically de Sitter geometries
essentially depends on the value of the mass of the field $m$: at some
treshold value of $m$ the non-oscillatory decay changes into
the oscillatory one.

The paper is organized as follows: in Sec. I we study, both analytically and
numerically, the late time behavior for spherically symmetrical
perturbations of Proca field in the background of Schwarzschild black
holes. Sec II. is devoted to the analytical consideration of late time
behavior for non-spherical perturbations. Finally, in Conclusion
we summarize the obtained results.

\section{Spherically symmetric perturbations}

We shall consider here the  Schwarzschild black hole solution with a
positive cosmological constant $\Lambda$, i.e. Schwarzschild and
Schwarzschild-de Sitter backgrounds in which the massive vector field
propagates. The black hole metric is given by
\begin{equation}
d s^2 = -h(r) d t^2 + h(r)^{-1} d r^2 + r^2 (d \theta^2 + \sin^2
\theta d \phi^2), \end{equation}
where
\begin{equation}
h(r) = 1-\frac{2 M}{r} - \frac{\Lambda r^2} {3} .
\end{equation}

For self-consistence, let us briefly deduce the wave equation
for spherically symmetric perturbations of the massive vector field
for general form of the spherically symmetric static metric (1).
The Proca equations have the form:
\begin{equation}
F^{\mu \nu}_{; \nu}- m^2 A^{\mu}=0, \quad F_{\mu \nu} = A_{\nu, \mu} -
A_{\mu, \nu}.
\end{equation}
These equation can be reduced in the spherically symmetric case to
the single equation of the form:
\begin{gather}
f(r) B_{, rr} - \frac{B_{, tt}}{f(r)} + \left(\frac{2 f(r)} {r} +
f'(r) \right) B_{, r} + \nonumber \\
\left(\frac{2 f'(r)}{r} - \frac{2 f(r)} {r^{2}} +  m^2 \right) B = 0,
\end{gather}
where
\begin{equation}
B = A_{r, t} - A_{t, r}.
\end{equation}

Introducing the usual ``tortoise'' radial coordinate $r_\star$ and a
new wave function
\begin{equation}
\Psi(r) = B(r) r \,\, ,
\end{equation}
the wave equation can be re-written in the form \cite{konoplya_sp1}:
\begin{equation}
\frac{\partial^2 \Psi(r_\star,t)}{\partial r_\star^{2}} - \frac{\partial^2
\Psi(r_\star,t)}{\partial t^{2}}  - V(r(r_\star)) \Psi (r_\star,t) = 0,
\end{equation}
with the effective potential
\begin{equation}
V(r) = \left(1-\frac{2 M}{r}  - \frac{\Lambda r^2} {3} \right)
\left(\frac{2}{r^2} - \frac{6 M} {r^3} + m^2 \right).
\label{Veff}
\end{equation}
Positive values of the cosmological term $\Lambda$ correspond to
asymptotically de Sitter solutions.

The time evolution of the massive vector field can be described by the
spectral decomposition method \cite{Leaver}. Thus,
\begin{gather}
\Psi(r_\star, t) =  \nonumber \\
= \int r_\star' (G (r_\star, r_\star'; t) \Psi_{t}(r_\star', 0)
+ G_{t} (r_\star, r_\star'; t) \Psi(r_\star', 0)).
\end{gather}
For $t > 0$, the Green function obeys the equation
\begin{gather}
\frac{\partial^2 G (r_\star, r_\star'; t)}{\partial r_\star^{2}} -
\frac{\partial^2 G (r_\star, r_\star'; t)}{\partial t^{2}}  -
V(r) G (r_\star, r_\star'; t) =
\nonumber \\
= \delta (t) \delta (r_\star - r_\star')
\end{gather}
In order to find the Fourier's component of the transformed Green
function  $G~ (r_\star, r_\star'; t)$, we need the two linearly
independent solutions $\Psi_{1}$, $\Psi_{2}$ to the homogeneous
equation
\begin{equation}
\frac{\partial^2 \Psi_{i}}{\partial r_\star^{2}} + (\omega^2
- V ) \Psi_{i} = 0, \quad i = 1, 2.
\end{equation}
If $\Lambda=0$, we require that the function $\Psi_{1}$ should
represent purely in-going waves at the event horizon ($r=r_{+}$):
$\Psi_{1} \sim e^{- i \omega r_\star}$ as $r_\star \rightarrow - \infty$.
The other function  $\Psi_{2}$ should damp exponentially at spatial
infinity:  $\Psi_{2} \sim e^{- \sqrt{\omega^2 - m^2} r_\star}$, as
$r_\star \rightarrow + \infty$. On the other hand, if  $\Lambda>0$,
the spacetime have a new feature, the cosmological horizon
($r=r_{c}$). In this case, the region of interest is the block
$r_{+}<r<r_{c}$. The effective potential decays exponentially to zero at both
limits, and therefore we require that $\Psi_{2} \sim e^{-
  i \omega r_\star}$, as $r_\star \rightarrow + \infty$.

To check the analytical results which will be found below, we used
a numerical characteristic integration scheme, based in the
light-cone variables $u = t - r_\star$ and $v = t + r_\star$. In the
characteristic initial value problem, initial data are specified on
the two null surfaces $u = u_{0}$ and $v = v_{0}$.
The discretization scheme applied, used for example in
\cite{Pullin,molina}, is
\begin{eqnarray}
\lefteqn{\Psi(N) = \Psi(W) + \Psi(E) -
\Psi(S) }  \nonumber \\
& & \mbox{} - \Delta^2 V(S) \frac{ \Psi(W) + \Psi(E)}{8} +
\mathcal{O}(\Delta^4)   \ ,
\label{d-uv-eq}
\end{eqnarray}
where we have used the definitions for the points: $N = (u + \Delta, v
+ \Delta)$, $W = (u + \Delta, v)$, $E = (u, v + \Delta)$ and $S =
(u,v)$.

In the asymptotically de Sitter scenario, we also employ a
semi-analytic approach to investigate the perturbative behavior of the
vector field. As will be seen, there is a region of the parameter
space of the geometry where the late-time perturbations are dominated by
quasinormal mode behavior. At this regime, we use the WKB formula
\begin{equation}
i \frac{\omega^2 - V_0}{\sqrt{-2V_0^{\prime\prime}}} - L_2 - L_3 - L_4 -
L_5 - L_6 = n + \frac{1}{2} \quad ,
\end{equation}
where $V_0$ is the height of the effective potential, and $V_0^{\prime\prime}$ is the second
derivative with respect to the tortoise coordinate of the potential at
the maximum. $L_2$, $L_3$  $L_4$, $L_5$ and $L_6$ are presented in
\cite{yer}.

\subsection{Schwarzschild black hole background}

\subsubsection{Intermediate late time behavior}

At intermediately late times, when $r \ll t \ll M/(M m)^{2}$, one can
neglect the effect of backscattering of the field from asymptotically
far regions and, following \cite{HodPiran1}, use the approximation $M
\ll r \ll  M/(Mm)^{2}$. One can expand the wave equation into power
series in $M/r$. Neglecting terms of order $\mathcal{O}((Mm/r)^{2})$,
we obtain
\begin{equation}
\frac{d^{2} \Psi}{d r^{2}} + \left(\omega^{2} - m^{2} + \frac{4 M
  \omega^{2} - 2 M m^{2}}{r} - \frac{2}{r^{2}}\right) \Psi = 0.
\end{equation}
Provided that the observer and the initial data are situated
far from the black hole we can further approximate
\begin{equation}
\frac{d^{2} \Psi}{d r^{2}} + \left(\omega^{2} - m^{2} -
\frac{2}{r^{2}}\right) \Psi = 0.
\end{equation}
The last equation is similar to Eq. (15) in \cite{HodPiran1},
therefore we can easily find the branch cut contribution to the Green function
for the above equation:
\begin{gather}
G^{C}(r_\star, r_\star', t) = \nonumber \\
\frac{8}{9 \pi A^{2}} \int_{0}^{m} \psi((r_\star, k)  \psi((r_\star',
k)   (k)^{3} e^{-i \omega t} d\omega
\end{gather}
where $k = \sqrt{m^2 - \omega^2}$, and $A$ is a normalization constant.

Note that the contribution to the Green function is dominated by
low-frequencies at late times. In the large $t$ limit the effective
contribution to the above integral is from $\sqrt{m^2 - \omega^2} =
\mathcal{O}(\sqrt{m/t})$. This is stipulated by rapidly oscillating
term $ e^{- i \omega t}$ so, that there is cancellation between the
positive and negative parts of the integrand. Finally, similar to
\cite{HodPiran1}, for $t \gg m^{-1}$ we obtain for a fixed radius, and
$r_\star, r_\star' \ll t$:
\begin{equation}
G^{C}(r_\star, r_\star', t) = \sqrt{\frac{2}{\pi}} \frac{m^{3/2} (r_\star
r_\star')^{2} t^{-5/2} \cos{(m t - (5 \pi/4))} } {6 !}
\end{equation}



\begin{figure}
\resizebox{\linewidth}{!}{%
\includegraphics*{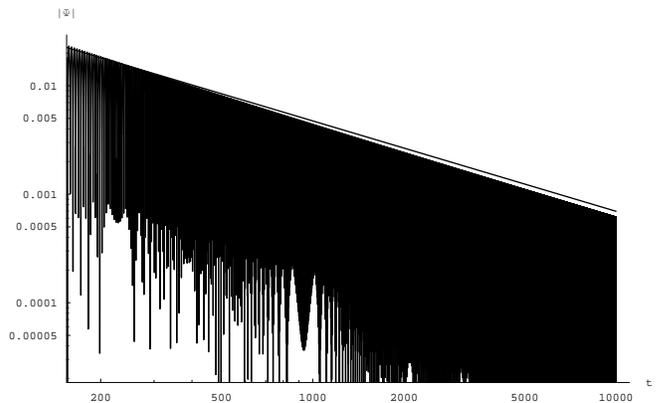}}
\caption{Late-time behavior of the massive vector field perturbation ($Mm=1$)
  in asymptotically flat geometry, at $r = 4.5M$ (initial conditions have been chosen constant $\Psi=1$ on the null cone).
  The analytically obtained law $\Psi\propto t^{-5/6}$ is shown by the line.
  At late times numerically we have the power coefficient $-0.844$, what differs from $-5/6$ only by $1.3\%$.}
\label{Wave-function-S1}
\end{figure}

\begin{figure}
\resizebox{\linewidth}{!}{%
\includegraphics*{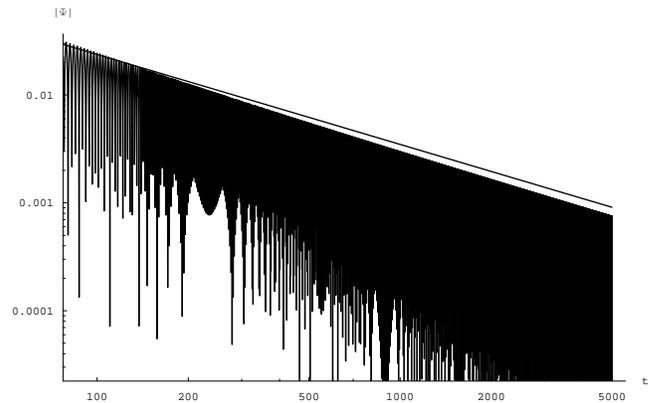}}
\caption{Late-time behavior of the massive vector field perturbation ($Mm=1.5$)
  in asymptotically flat geometry, at $r = 1.6M$ (initial conditions have been chosen constant $\Psi=1$ on the null cone).
  The analytically obtained law $\Psi\propto t^{-5/6}$ is shown by the line.
  At late times numerically we have the power coefficient $-0.849$, what differs from $-5/6$ only by $1.76\%$.}
\label{Wave-function-S2}
\end{figure}


Therefore one can conclude that at intermediate times the field $\Psi$
decays at a fixed radius according to the law
\begin{equation}
\Psi \sim t^{-5/2} \sin{mt},
\end{equation}
and at the black hole event horizon, the decay is the same
\begin{equation}
\Psi \sim v^{-5/2} \sin{mt}.
\end{equation}
This analytically obtained result in the regime of intermediately
late times is confirmed by numerical characteristic integration with high accuracy,
what is illustrated in Figs. \ref{Wave-function-S1} and \ref{Wave-function-S2}.
There one can see that at sufficiently late times the numerical envelope approaches the analytical law $\Psi\propto t^{-5/6}$.
The difference between numerical envelope and the line $t^{-5/6}$ is decreasing as time increases and is about 1\% for $t\sim 10000$ (see Fig. \ref{Wave-function-S1}).

\subsubsection{Asymptotic late-time behavior}

\begin{figure}
\resizebox{\linewidth}{!}{\includegraphics*{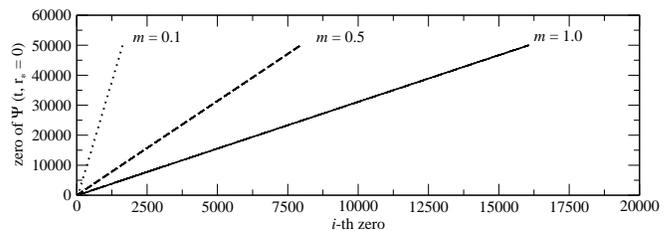}}
\caption{Numerical value of $t$ in which the wave function  is
  zero. Strait lines imply that period of oscillations is a constant. For late
  times, the lines $t_{zero} (i) = 2\pi i / m$ are excellent
  approximations of the numerical values.}
\label{Zeros-S}
\end{figure}

In \cite{koyama1}, it was shown that at asymptotically late times
the law of decay is stipulated by the behavior of the wave equation
at large $r$. Thus, in the region $r/M \gg 1$, expanding the wave
equation in powers of $M/r$, we obtain
\begin{equation}
\frac{d^{2} \psi}{d x^{2}} + \left(\frac{1}{4} +\frac{a}{x}-
\frac{b^{2} - 1/4}{x^{2}}\right) \psi = 0,
\end{equation}
where we used a new variable $x = 2 r \sqrt{m^2 - \omega^2}$, and
$a$, $b$ are constants depending on $m$, $M$, and $\omega$.
Following the papers  \cite{koyama2} and  \cite{koyama1}, one can show
that the monopole massive vector field at asymptotically late times $t
\rightarrow \infty $ undergoes oscillatory inverse power law decay
\begin{equation}
\Psi \sim t^{-5/6} \sin{m t}, \quad  t \rightarrow \infty.
\end{equation}
This power-law envelope in shown in  Figs. \ref{Wave-function-S1} and
\ref{Wave-function-S2}. In the range $m t \gg 1/(m M)^{2}$, the
smaller the value $m M$ is,  the later the  $t^{-5/6}$ tail begins to
dominate. In the range $m t \gg m M$, on the contrary, the larger the
value $M m$ is, the later $t^{-5/6}$ tail begins to dominate. The
numerical investigation show that, both in the intermediate and late
time regime, the oscillation period can be perfectly approximated by
\begin{equation}
\textrm{period} = \frac{2 \pi}{m} .
\end{equation}
This is illustrated in Fig. \ref{Zeros-S}.

\subsection{Schwarzschild-de Sitter black hole background}

The introduction of a positive cosmological constant changes
drastically the late-time tails. Direct numerical
integration shows two distinct late-time decay regimes. We have
observed that the qualitative aspects of the field decay are
independent of the value of the cosmological constant, in the
non-extreme regime.

For fields with small values of mass, we observe that the late-time
decay is exponential and non-oscillatory:
\begin{equation}
\Psi \sim e^{k_{1} t} \, .
\end{equation}
The exponential index $k_{1}$ approaches zero as the
field mass tends to zero. This decay mode is similar to the tail
observed in the massless vector perturbation in Schwarzschild-de
Sitter black hole backgrounds.

For $m$ greater than a critical value, the decay is dominated by
quasinormal modes. In this regime, it is possible to use the well
known semi-analytical WKB approach to calculate the quasinormal
frequencies. In this approach, by the way, the convergence is very
good. We present some results in table \ref{mqn-sc-cfm}, where it is
compared frequencies calculated directly from the characteristic
integration and first and sixth order WKB formulas.

\begin{table*}[tp]
\label{mqn-sc-cfm}
\caption{Fundamental quasinormal frequencies for the massive vector field
  perturbation around the SdS black hole for several values of the
  parameters $m$ and $\Lambda$. The black hole mass is set to $M=1$.}
\begin{ruledtabular}
\begin{tabular}{cccccccc}
\multicolumn{2}{l}{}                           &
\multicolumn{2}{c}{Direct Integration}         &
\multicolumn{2}{c}{WKB-$1^{th}$ order}                    &
\multicolumn{2}{c}{WKB-$6^{th}$ order}                \\ \\
$\Lambda$         & $m$        &
Re($\omega_0$) & -Im($\omega_0$) &
Re($\omega_0$) & -Im($\omega_0$) &
Re($\omega_0$) & -Im($\omega_0$) \\ \\
\hline \\
$10^{-4}$ & 0.5  & 0.476680  & $4.506081 \times 10^{-3}$ &
                   0.4771778 & $4.582987 \times 10^{-3}$ &
                   0.4773061 & $4.569051 \times 10^{-3}$ \\
$10^{-4}$ & 1.0  & 0.950732  & $4.764696 \times 10^{-3}$ &
                   0.9514156 & $4.717815 \times 10^{-3}$ &
                   0.9514795 & $4.715673 \times 10^{-3}$  \\
$10^{-4}$ & 2.0  & 1.900517 & $4.794143 \times 10^{-3}$ &
                   1.901453 & $4.744193 \times 10^{-3}$ &
                   1.901485 & $4.743740 \times 10^{-3}$ \\

$10^{-4}$ & 5.0  & 4.751232 & $4.873461 \times 10^{-3}$ &
                   4.752683 & $4.751188 \times 10^{-3}$ &
                   4.752695 & $4.751119 \times 10^{-3}$ \\
$10^{-4}$ & 10.0 & 9.586787 & $5.166647 \times 10^{-3}$ &
                   9.505094 & $4.752174 \times 10^{-3}$ &
                   9.505101 & $4.752157 \times 10^{-3}$ \\ \\
$10^{-3}$ & 0.5  & 0.453757   & $1.368941 \times 10^{-2}$
                 & 0.4526006  & $1.368657 \times 10^{-2}$
                 & 0.4534351  & $1.333422 \times 10^{-2}$  \\
$10^{-3}$ & 1.0  & 0.893654   & $1.397031 \times 10^{-2}$
                 & 0.8935073  & $1.397114 \times 10^{-2}$
                 & 0.89364734 & $1.396942 \times 10^{-2}$ \\
$10^{-3}$ & 2.0  & 1.781756   & $1.404640 \times 10^{-2}$
                 & 1.781636   & $1.404642 \times 10^{-2}$
                 & 1.781694   & $1.404750 \times 10^{-2}$  \\
$10^{-3}$ & 5.0  & 4.451429   & $1.405823 \times 10^{-2}$
                 & 4.450396   & $1.406723 \times 10^{-2}$
                 & 4.450418   & $1.406745 \times 10^{-2}$ \\
$10^{-3}$ & 10.0 & 8.902677   & $1.405555 \times 10^{-2}$
                 & 8.899741   & $1.407019 \times 10^{-2}$
                 & 8.899752   & $1.407024 \times 10^{-2}$ \\ \\
$10^{-2}$ & 0.5  & 0.390560   & $4.126141 \times 10^{-2}$
                 & 0.3919454  & $3.917318 \times 10^{-2}$
                 & 0.3905636  & $4.122846 \times 10^{-2}$  \\
$10^{-2}$ & 1.0  & 0.751753   & $3.777082 \times 10^{-2}$
                 & 0.7529941  & $3.750283 \times 10^{-2}$
                 & 0.7522548  & $3.772474 \times 10^{-2}$  \\
$10^{-2}$ & 2.0  & 1.489152   & $3.739187 \times 10^{-2}$
                 & 1.490789   & $3.722330 \times 10^{-2}$
                 & 1.490332   & $3.727043 \times 10^{-2}$  \\
$10^{-2}$ & 5.0  & 3.713665   & $3.725257 \times 10^{-2}$
                 & 3.716378   & $3.715591 \times 10^{-2}$
                 & 3.716186   & $3.716325 \times 10^{-2}$  \\
$10^{-2}$ & 10.0 & 7.427974   & $3.731987 \times 10^{-2}$
                 & 7.429733   & $3.714668 \times 10^{-2}$
                 & 7.429636   & $3.714851 \times 10^{-2}$  \\ \\
\end{tabular}
\end{ruledtabular}
\end{table*}

\begin{figure}
\resizebox{\linewidth}{!}{\includegraphics*{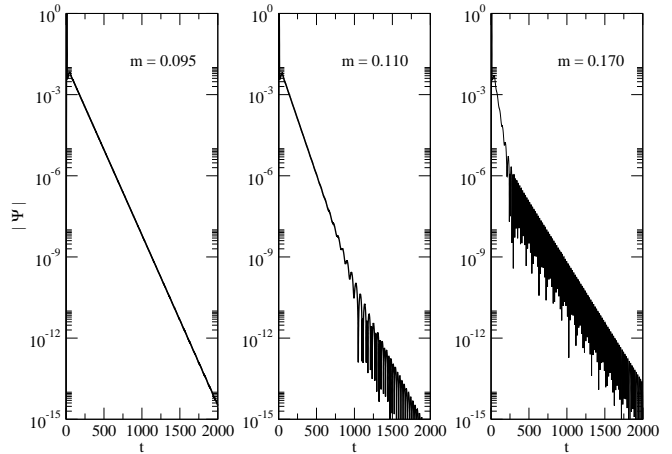}}
\caption{Semi-log graphs of the massive vector field perturbation in de
  Sitter geometries, at $r_\star = 0$. Increasing the mass, it is
  observed the transition from non-oscillatory to oscillatory
  asymptotic decay. In the graphs, $M=1$ and $\Lambda = 10^{-4}$. }
\label{dS-waves}
\end{figure}

\begin{figure}
\resizebox{\linewidth}{!}{\includegraphics*{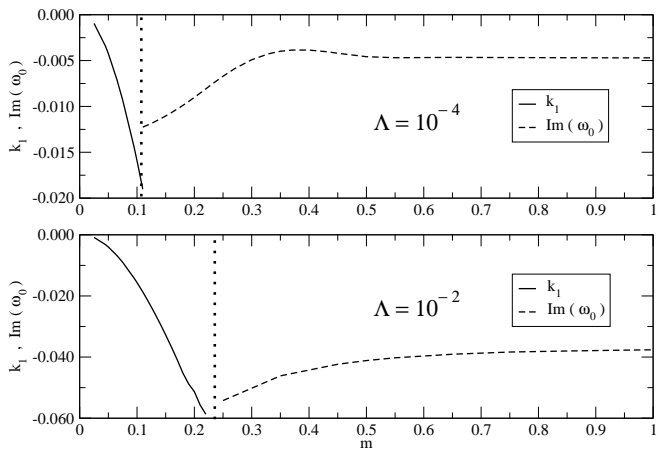}}
\caption{Graphs of $k_{1}$ and $k_{2}$ with the mass $m$. The vertical
 dotted lines approximate the masses where it is observed the
 transition from non-oscillatory to oscillatory
  asymptotic decay. In the graphs $M=1$. }
\label{regions}
\end{figure}

The transition from non-oscillatory to  oscillatory decay is shown in
Fig. \ref{dS-waves}. We also observe that, the larger the cosmological
constant, the greater is the critical mass $m$ where the oscillatory
regime dominates. For high values of $m$, the exponential envelope
index $k_{2}$ approaches a constant value. These observations are
illustrated in Fig. \ref{regions}.

One interesting limit of the Schwarzschild-de Sitter black hole
geometry is its near extreme limit, when the  event and cosmological
horizons are very close, and the cosmological constant approach its
maximum value ($1/9 M^2$).

\begin{figure}
\resizebox{1\linewidth}{!}{\includegraphics*{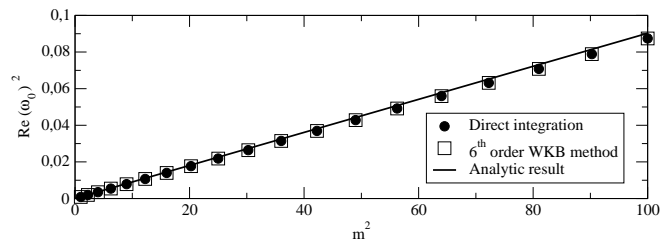}}
\caption{Graph of $\textrm{Re} (\omega_{0})^2$ as a function of $m^{2}$, in the near-extreme limit positive $\Lambda$
    limit. The bullets are the values calculated from the
    time-evolution profiles, squares are values obtained from WKB
    formula, and the strait line represents the results
    from the expression (\ref{QNM_near_extreme}). The parameters in
    this graph are $\Lambda= 0.11082$ and $M=1$. The
    differences between the results are under 1\%.}
\label{near_extreme}
\end{figure}

In the near extreme limit, an analytic
expression can be obtained for the quasinormal frequencies, which
dominate the late-time decay if the mass $m$ is large
enough. Following the work presented in \cite{Cardoso-03,Molina-03}, we have
\begin{gather}
\textrm{Re} (\omega_{n}) = \frac{\Lambda (1 - 9 M^2 \Lambda)}{3}
\sqrt{\frac{m^2}{\Lambda} - \frac{1}{4} } \,\, , \nonumber \\
\textrm{Im} (\omega_{n}) = - \frac{\Lambda (1 - 9 M^2 \Lambda)}{3}
\left( n + \frac{1}{2} \right) \,\, , \nonumber \\
n = 0, 1, 2, \ldots \quad .
\label{QNM_near_extreme}
\end{gather}
The above formula is very well confirmed numerically:
Fig.\ref{near_extreme} shows a comparison of the values obtained by
direct numerical integration, WKB formula and from
Eq. (\ref{QNM_near_extreme}).

\section{Perturbations of higher multipoles}

In this section we shall consider only the case of Schwarzschild black
hole geometry. It is known from the work of Galtsov, Pomerantseva and Chizhov
\cite{Galtsov} that the perturbation equations for massive vector
field in the Schwarzschild background cannot be decoupled for general
case of arbitrary multipolarity $\ell$. Yet, far from black hole the
equations can be decoupled \cite{Galtsov}, what allows us judge about
asymptotically late-time behavior.  As in the paper
\cite{Galtsov} the perturbation equations are not shown in explicit
form, but rather as an operator equations, we shall have to repeat
here some of the derivations of \cite{Galtsov}, in order to proceed
our analysis of tail behavior.

The Proca equation reads
\begin{equation}
F^{\mu \nu}_{; \nu}- m^2 A^{\mu}=0, \quad F^{\mu \nu} = A_{\nu, \mu} -
A_{\mu, \nu} .
\end{equation}
Using the Newman-Penrose formalism, we go over to the tetrad
projections of the field $F_{\mu \nu}$
\begin{equation}
\Phi_{0} = F_{\mu \nu} l^{\mu} m^{\nu}, \quad \Phi_{1} = \frac{1}{2}
F_{\mu \nu} ( l^{\mu} n^{\nu} +  m^{*\mu} m^{\nu}),
\end{equation}
\begin{equation}
\Phi_{2} =  F_{\mu \nu} m^{*\mu} n^{\nu},
\end{equation}
and of the vector-potential
\begin{equation}
A^{\mu} = A_{l}n^{\mu} + A_{n}l^{\mu} - A_{m}m^{*\mu} - A_{m*}m^{\mu}.
\end{equation}
After projecting of the equations (15) onto the tetrad's vectors and
expansions of $\Phi_{i}$ and $A_{\alpha}$ into spin-weighted spherical
harmonics \cite{Chandra}
\begin{equation}
\Phi_{0} = \sum_{\ell = 1, |m| \leq \ell}^{\infty} e^{- i \omega t + i m
\phi} (_{1}Y_{\ell m}) \Psi_{0}^{\ell m}(r),
\end{equation}
\begin{equation}
\Phi_{1} = \sum_{\ell = 0, |m| \leq \ell}^{\infty} e^{- i \omega t + i m
\phi} (_{0}Y_{\ell m}) \Psi_{1}^{\ell m}(r),
\end{equation}
\begin{equation}
\Phi_{2} = \sum_{\ell = 1, |m| \leq \ell}^{\infty} e^{- i \omega t + i m
\phi} (_{-1}Y_{\ell m}) \Psi_{2}^{\ell m}(r),
\end{equation}
\begin{equation}
A_{m} = \sum_{\ell = 1, |m| \leq \ell}^{\infty} e^{- i \omega t + i m
\phi} (_{1}Y_{\ell m}) \Psi_{3}^{\ell m}(r),
\end{equation}
\begin{equation}
A_{m^{*}} = \sum_{\ell = 1, |m| \leq \ell}^{\infty} e^{- i \omega t + i m
\phi} (_{-1}Y_{\ell m}) \Psi_{4}^{\ell m}(r),
\end{equation}
\begin{equation}
A_{n} = \sum_{\ell = 0, |m| \leq \ell}^{\infty} e^{- i \omega t + i m
\phi} (_{0}Y_{\ell m}) \Psi_{5}^{\ell m}(r),
\end{equation}
\begin{equation}
A_{l} = \sum_{\ell = 0, |m| \leq \ell}^{\infty} e^{- i \omega t + i m
\phi} (_{0}Y_{\ell m}) \Psi_{6}^{\ell m}(r),
\end{equation}
one can exclude the functions $\Psi_{3}$, $\Psi_{4}$,  $\Psi_{5}$, $\Psi_{6}$
and obtain the three equations for the three scalar
functions $\Psi_{0}$,  $\Psi_{1}$, and $\Psi_{2}$  \cite{Galtsov}:
\begin{equation}
(D_{0} D_{0}^{+} - U) \Psi_{0} + \sqrt{2} \lambda r^{-2} \Psi_{1} = 0,
\end{equation}

\begin{equation}
(D_{0}^{+} D_{0} - U) \Psi_{2} + \sqrt{2} \lambda r^{-2} \Psi_{1} = 0,
\end{equation}

\begin{gather}
r^{3} ( D_{1}^{+} r^{-4} D_{0} r + D_{1} r^{-4} D_{0}^{+} r )\Psi_{1}
- 2 U \Psi_{1} + \nonumber \\
+ 2 \sqrt{2} (r^{2} - 2 M r)^{-1} (\Psi_{0} + \Psi_{2}) = 0 .
\end{gather}
Here we used the operators $D_{j}$, $j=0$, $1$:
\begin{equation}
D_{j} = \partial_{r} - \frac{i \omega r}{r - 2 M}+ j \frac{2(r- M)}{r (r
- 2 M)}
\end{equation}
\begin{equation}
D_{j}^{+} = \partial_{r} - \frac{i \omega r}{r - 2 M}+ j \frac{2(r- M)}{r (r
- 2 M)}.
\end{equation}
Here $\lambda = \ell (\ell + 1)$, and
$$U = (\lambda^{2} + \mu^{2} r^{2})(r^{2}- 2 M r)^{-1}$$.

Using  new notations:
\begin{equation}
x = \frac{r}{2 M} -1, \quad \mu = 2 \sqrt{2} m M, \quad w = 2 M \omega
\end{equation}
we obtain the system of equations

\begin{widetext}
\begin{equation}
\Psi_{0, xx} +
\left[
w^{2} \left(1+ \frac{1}{x} \right)^{2} - \frac{\mu^{2}}{2} \left(1+
\frac{1}{x} \right)
- \frac{i w}{x^{2}} - \frac{\lambda^{2}}{x(x+1)}\right] \Psi_{0} +
\frac{\sqrt{2} \lambda}{x (x + 1)}  \Psi_{1} = 0
\label{Psi_0xx}
\end{equation}
\begin{equation}
\Psi_{2, xx} +
\left[
w^{2} \left(1+ \frac{1}{x} \right)^{2} - \frac{\mu^{2}}{2} \left(1+
\frac{1}{x} \right)
+ \frac{i w}{x^{2}} - \frac{\lambda^{2}}{x(x+1)}\right]  \Psi_{2} +
\frac{\sqrt{2} \lambda}{x (x + 1)} \Psi_{1}  = 0
\label{Psi_2xx}
\end{equation}
\begin{gather}
\Psi_{1, xx} + \frac{1}{x ( x + 1 )} \Psi_{1, x} +
\left[
w^{2} \left(1+ \frac{1}{x} \right)^{2} - \frac{\mu^{2}}{2} \left(1+
\frac{1}{x} \right)
+ \frac{x - 2 x^{2}}{x^{2} (x+1)^{2}} - \frac{\lambda^{2}}{x(x+1)}\right] \Psi_{1}
\nonumber \\
 +\frac{\sqrt{2} \lambda}{x (x + 1)} (\Psi_{0} +\Psi_{2}) = 0.
\label{Psi_1xx}
\end{gather}

\end{widetext}

To check these equations one can consider here spherically symmetric
perturbations, for which $\Psi_{0}= \Psi_{2} = 0$,  while $\Psi_{1}$ is non-zero
and obeys the wave equation (\ref{Psi_1xx})  with $\ell =
0$. After coming back to $r$-coordinate, Eq.(\ref{Psi_1xx}) reduce to
Eq. (\ref{Veff}).

\subsection{Intermediate late-time behavior}

Now, let us consider phenomenologically interesting region  $m M \ll
1$, what corresponds to the notion of ``small'' perturbations, i.e.
perturbations where the mass of the field $m$ is much smaller then the
mass of the black hole. Again, at intermediately late times
($r \ll t \ll M/(M m)^{2}$), we  neglect the effect of backscattering
of the field from asymptotically far region, and use the
approximation $M \ll r \ll  M/(M m)^{2}$. For an observer and an initial
data situated far from the black hole,  from the above
Eqs. (\ref{Psi_0xx})-(\ref{Psi_1xx}), we obtain the following equations
\begin{equation}
\Psi_{0, xx} +
\left(
w^{2}  - \frac{\mu^{2}}{2}  - \frac{\lambda^{2}}{x^{2}}\right) \Psi_{0} +
\frac{\sqrt{2} \lambda}{x^{2}}  \Psi_{1} = 0
\end{equation}
\begin{equation}
\Psi_{2, xx} +
\left(
w^{2}  - \frac{\mu^{2}}{2}  - \frac{\lambda^{2}}{x^{2}}\right) \Psi_{2} +
\frac{\sqrt{2} \lambda}{x^{2}}  \Psi_{1} = 0
\end{equation}
\begin{equation}
\Psi_{1, xx} +
\left(w^{2}  - \frac{\mu^{2}}{2}  - \frac{\lambda^{2} + 2}{x^{2}}\right)
\Psi_{1}
 +\frac{\sqrt{2} \lambda}{x^{2}} (\Psi_{0} +\Psi_{2}) = 0.
\end{equation}

The above equations can be easily diogonalized by linear
$x$-independent transformations of the $\Psi_{i}$-functions.
\begin{equation}
\tilde{\Psi}_{0, xx} +
\left(
w^{2}  - \frac{\mu^{2}}{2}  - \frac{\lambda^{2}}{x^{2}}\right) \tilde{\Psi}_{0} = 0
\end{equation}
\begin{equation}
\tilde{\Psi}_{2, xx} +
\left(
w^{2}  - \frac{\mu^{2}}{2}  - \frac{1 + \lambda^{2} - \sqrt{
1+ 4 \lambda^{2}}}{x^{2}}\right) \tilde{\Psi}_{2} = 0
\end{equation}
\begin{equation}
\tilde{\Psi}_{1, xx} +
\left(
w^{2}  - \frac{\mu^{2}}{2}  - \frac{1 + \lambda^{2} + \sqrt{
1+ 4 \lambda^{2}}}{x^{2}}\right) \tilde{\Psi}_{1} = 0
\end{equation}

The above equations can be
re-written in the following way:
\begin{equation}
\tilde{\Psi}_{0, xx} +
\left[
w^{2}  - \frac{\mu^{2}}{2}  - \frac{\ell (\ell + 1)}{x^{2}}\right]
\tilde{\Psi}_{0} = 0, \quad \ell = 1, 2, 3,  \ldots
\label{psi_hat_0}
\end{equation}
\begin{equation}
\tilde{\Psi}_{2, xx} +
\left[
w^{2}  - \frac{\mu^{2}}{2}  - \frac{\ell (\ell - 1)}{x^{2}}\right]
\tilde{\Psi}_{2} = 0, \quad \ell = 1, 2, 3, \ldots
\label{psi_hat_1}
\end{equation}
\begin{equation}
\tilde{\Psi}_{1, xx} +
\left[
w^{2}  - \frac{\mu^{2}}{2}  - \frac{\ell (\ell + 3) + 2
}{x^{2}}\right] \tilde{\Psi}_{1} = 0, \quad \ell = 0,1,2,\ldots
\end{equation}
One can see that $\ell (\ell + 1)$ in Eq.(\ref{psi_hat_0}) runs values
$2,6,12,20,\ldots$,  $\ell (\ell - 1)$ in Eq.(\ref{psi_hat_1}) runs
the values $0,2,6,12,20,\ldots $, while $\ell (\ell + 3) + 2$ runs
values $2,6,12,20,\ldots$ . Therefore we can finally re-write the
equations in the following unique  form:
\begin{equation}
\tilde{\Psi}_{j, xx} +
\left(
w^{2}  - \frac{\mu^{2}}{2}  - \frac{\ell_{j} (\ell_{j} +
1)}{x^{2}}\right) \tilde{\Psi}_{j} = 0,
\end{equation}
where $ j = 0, 1, 2.$, and
\begin{eqnarray}
\ell_{0} & = & \ell,  \quad \ell = 1,2,\ldots  \\
\ell_{1} & = & \ell + 1, \quad  \ell = 0,1,2,\ldots \\
\ell_{2} & = & \ell - 1, \quad  \ell = 1,2,\ldots
\end{eqnarray}

Repeating analysis of the section II, we can easily see that the
intermediate late-time behavior will be
\begin{equation}
\tilde{\Psi}_{0} \sim t^{-(\ell + 3/2)} \sin{mt}, \quad \ell = 1,2,\ldots
\end{equation}
\begin{equation}
\tilde{\Psi}_{1} \sim t^{-(\ell + 5/2)} \sin{mt} \quad \ell = 0,1,2,\ldots
\end{equation}
\begin{equation}
\tilde{\Psi}_{2} \sim t^{-(\ell + 1/2)} \sin{mt} \quad \ell = 1,2,\ldots
\end{equation}

\subsection{Asymptotic late-time behavior}

If in the regime $ x\gg1 $, we  take into consideration  sub-dominant terms of order
$ \omega/x $ in Eq. (\ref{Psi_0xx})-(\ref{Psi_1xx}), we come to
to the following equations:
\begin{widetext}
\begin{equation}
\Psi_{0, xx} +
\left[
w^{2} \left(1+ \frac{2}{x} \right) - \frac{\mu^{2}}{2} \left(1+
\frac{1}{x} \right) - \frac{\lambda^{2}}{x^{2}}\right] \Psi_{0} +
\frac{\sqrt{2} \lambda}{x^{2}}  \Psi_{1} = 0
\label{A}
\end{equation}
\begin{equation}
\Psi_{2, xx} +
\left[
w^{2} \left(1+ \frac{2}{x} \right) - \frac{\mu^{2}}{2} \left(1+
\frac{1}{x} \right) - \frac{\lambda^{2}}{x^{2}}\right]  \Psi_{2} +
\frac{\sqrt{2} \lambda}{x^{2}} \Psi_{1}  = 0
\label{B}
\end{equation}
\begin{gather}
\Psi_{1, xx} +
\left[ w^{2} \left(1+ \frac{2}{x} \right) - \frac{\mu^{2}}{2} \left(1+
\frac{1}{x} \right)  - \frac{\lambda^{2}+ 2}{x^{2}}\right] \Psi_{1} +\frac{\sqrt{2} \lambda}{x^{2}} (\Psi_{0} +\Psi_{2}) = 0.
\label{C}
\end{gather}

\end{widetext}

Following \cite{Galtsov}, one can introduce a new variable $z = 2
\sqrt{m^{2} - \omega^{2}}$ r, then one can see that the system of
equations (\ref{A}-\ref{C}) has a solution \cite{Galtsov}:
\begin{equation}
\tilde{\Psi}_{j} = C_{j} W_{p, q} (z), \quad j = 0, 1, 2.
\end{equation}
where  $W_{p, q}$ is the Whittaker function \cite{special_functions}
\begin{equation}
q = (2 \omega^{2} - m^{2}) M/\sqrt{m^{2} -\omega^{2}},
\end{equation}
and for $p$, as was shown in \cite{Galtsov}, one has the following system of equations
\begin{equation}
(p^{2} -\lambda^{2} - (1/4)) C_{0} + \sqrt{2} \lambda C_{1} = 0,
\end{equation}
 \begin{equation}
(p^{2} -\lambda^{2} - (1/4)) C_{2} + \sqrt{2} \lambda C_{1} = 0,
\end{equation}
\begin{equation}
(p^{2} -\lambda^{2} - (9/4)) C_{1} + \sqrt{2} \lambda (C_{2} + C_{0}) = 0.
\end{equation}
Therefore, one immediately has values of $p$:
\begin{equation}
p = \ell + (1/2) + \sigma, \quad \sigma = 0, \pm1.
\end{equation}

As was shown in \cite{koyama1} for massive scalar field,
the asymptotically late-time behavior of the perturbations governed by
the wave equation which looks at asymptotically large $r$ as a
Whittaker equation, the decay law is universal and does not depend
on $ \ell $. Thus, in a similar fashion with \cite{koyama1}, we find
that the decay at asymptotically late times $ t \rightarrow \infty $
again is
\begin{equation}
\tilde{\Psi}_{j} \sim t^{-5/6} \sin{mt}, \quad j = 0,1,2.
\end{equation}

\section{Conclusions}

In the present paper we have considered evolution of perturbations of
massive vector field at late times. For spherically symmetric case,
the perturbation equations can be reduced to a single wave-like
equation, and therefore can be analyzed numerically with
characteristic integration method.  In this case, time-domain picture
of a signal has good agreement with analytical predictions for the
considered range of times.

General vector field perturbations
can be represented by three scalar functions $\Psi_{i}$, ($i = 0, 1,
2$). At intermediate late times  $r \ll t \ll M/(M m)^{2}$, these
three functions decay according to different laws which depend on a
multipole number $\ell$, $\tilde{\Psi}_{0} \sim t^{-(\ell + 3/2)}$,
$\tilde{\Psi}_{1} \sim t^{-(\ell + 5/2)}$, $\tilde{\Psi}_{2}
\sim t^{-(\ell + 1/2)}$.
On the contrary, at asymptotically late times $t \rightarrow \infty $,
the decay law is universal, i.e. the same for all three functions
and does not depend on $\ell$: $\tilde{\Psi}_{j} \sim t^{-5/6} \sin{mt}$.
Note again, that the same asymptotically late-time decay low was previously
found in \cite{jing}, \cite{koyama1} for scalar and spinor (Dirac)
perturbations around Schwarzschild black hole. Summarizing this
we can conclude that at asymptotically late times  $t \rightarrow
\infty $, the decay low is {\it universal} and \emph{does not depend
  on spin of a field $s$ and multipole number  $\ell$}:
\begin{equation}
\Psi_{s} \sim t^{-5/6} \sin{mt}, \quad s = 0, 1/2, 1.
\end{equation}

The latter law is obtained analytically and confirmed numerically with high accuracy.

For Schwarzschild-de Sitter black hole
it is observed quite non-trivial tail behavior. In the considered
asymptotically de Sitter geometries, the power-law oscillatory tails
are replaced by purely exponential tails, if $m$ is small, or
by oscillatory quasinormal decay, if the parameter $m$ is large enough.
Although there is some universality for the tail behavior in the
context of Schwarzschild geometry, this is not so in
non-asymptotically flat backgrounds.

An interesting question, which was beyond our study, is whether the
asymptotically flat background universality of tail behavior keeps in
higher dimensions, and for a more general then Schwarzschild backgrounds?
Yet, for Kerr black hole there is no hope to separate variables in the
perturbed Proca equations, while for a Reissner-Nordstr\"{o}m black hole
such a separation is certainly possible for spherically symmetric
perturbations, and charged Proca field should be considered
instead. We hope that future investigations  will clarify all these
points.


\begin{acknowledgments}
 This work was partially supported by \emph{Funda\c{c}\~{a}o de Amparo
\`{a} Pesquisa do Estado de S\~{a}o Paulo (FAPESP)} and \emph{Conselho
 Nacional de Desenvolvimento Cient\'ifico e Tecnol\'ogico (CNPq)}, Brazil.
\end{acknowledgments}


\end{document}